\documentclass[journal]{IEEEtran}

\ifCLASSINFOpdf
\else
\usepackage[dvips]{graphicx}
\fi
\usepackage{url}

\usepackage{graphicx}

\usepackage{amsfonts, amsmath, amsthm, bm, amssymb}
\usepackage{multirow}

\def \w {\mathbf{w}}

\def \rr {\mathbf{r}} 
\usepackage{booktabs}

\usepackage[colorlinks,
linkcolor=blue,
anchorcolor=green,
urlcolor=black,
citecolor=blue
]{hyperref}

\usepackage{caption}

\begin{document}
	
\title{WaveNet-Volterra Neural Networks for Active Noise Control: A Fully Causal Approach}

\author{Lu Bai, Mengtong Li, Siyuan Lian, \IEEEmembership{Member, IEEE}, Kai Chen, \IEEEmembership{Member, IEEE}, and Jing Lu, \IEEEmembership{Member, IEEE}

\thanks{Manuscript received xxxxxxx xx, 2025; revised xxxxxxx xx, 2025; accepted xxxxxxx xx, 2025. Date of publication xxx xx, 2025; date of current version xxx xx, 2025. The associate editor coordinating the review of this manuscript and approving it for publication was xxx. This work was supported by NSF China No. 12274221 and the AI and AI for Science Project of Nanjing University. (Corresponding author: Jing Lu.)}
\thanks{Lu Bai, Mengtong Li, Siyuan Lian, Kai Chen and Jing Lu are with the Key Laboratory of Modern Acoustics, Institute of Acoustics, Nanjing University, Nanjing, 210093, China, and also with NJU-Horizon Intelligent Audio Lab, Nanjing Institute of Advanced Artificial Intelligence, Nanjing 210014, China (e-mail: lubai@smail.nju.edu.cn; mengtong.li@smail.nju.edu.cn; siyuan.lian@smail.nju.edu.cn; chenkai@nju.edu.cn; lujing@nju.edu.cn).}}
		
\markboth{Journal of \LaTeX\ Class Files, Vol. 14, No. 8, August 2015}
{Shell \MakeLowercase{\textit{et al.}}: Bare Demo of IEEEtran.cls for IEEE Journals}
\maketitle
		
\begin{abstract}

Active Noise Control (ANC) systems are challenged by nonlinear distortions, which degrade the performance of traditional adaptive filters. While deep learning-based ANC algorithms have emerged to address nonlinearity, existing approaches often overlook critical limitations: (1) end-to-end Deep Neural Network (DNN) models frequently violate causality constraints inherent to real-time ANC applications; (2) many studies compare DNN-based methods against simplified or low-order adaptive filters rather than fully optimized high-order counterparts. In this letter, we propose a causality-preserving time-domain ANC framework that synergizes WaveNet with Volterra Neural Networks (VNNs), explicitly addressing system nonlinearity while ensuring strict causal operation. Unlike prior DNN-based approaches, our method is benchmarked against both state-of-the-art deep learning architectures and rigorously optimized high-order adaptive filters, including Wiener solutions. Simulations demonstrate that the proposed framework achieves superior performance over existing DNN methods and traditional algorithms, revealing that prior claims of DNN superiority stem from incomplete comparisons with suboptimal traditional baselines. Source code is available at https://github.com/Lu-Baihh/WaveNet-VNNs-for-ANC.git.

\end{abstract}


\IEEEpeerreviewmaketitle

\section{Introduction}
		
\IEEEPARstart{A}{ctive} noise control (ANC) reduces noise by generating an anti-noise signal of equal amplitude and opposite phase \cite{kuo1999active}. Compared to passive noise control, ANC is particularly effective for low-frequency noise and has been widely applied in automobiles \cite{cheer2012active}, headphones \cite{1597204}, and aircraft \cite{ELLIOT1990219}, among other fields \cite{wang2022improving,yang2023active,li2023augmented}.
		
Traditional ANC is implemented using adaptive filters that optimize filter weights by minimizing the error signal. The Wiener filter is theoretically optimal in the least-squares sense, but its reliance on statistical parameters limits its performance in time-varying environments \cite{elliott2000signal}. The filtered-reference least mean square (FxLMS) algorithm \cite{elliott1987multiple} and its extensions  are widely used due to their simplicity and robustness \cite{li2023augmented,li2023distributed}. 
The frequency domain filtered-reference least mean square (FD-FxLMS) algorithm \cite{yang2018frequency} efficiently performs linear convolution and correlation calculations in the frequency domain by using the discrete Fourier transform (DFT), which can improve the convergence rate and ameliorate the instability of the time domain filtered-reference least mean square (TD-FxLMS) algorithm with respect to impulsive signals \cite{zhang2019normalized}. Recent advances, such as the online decoupling and whitening frequency-domain filtered-error LMS (ODW-FDFeLMS) algorithm, have further improved convergence in multi-channel systems \cite{lian2024online}.
		
\begin{figure}[!t]
	\vspace{-0.1cm}
	\centering
	\includegraphics[width=\linewidth]{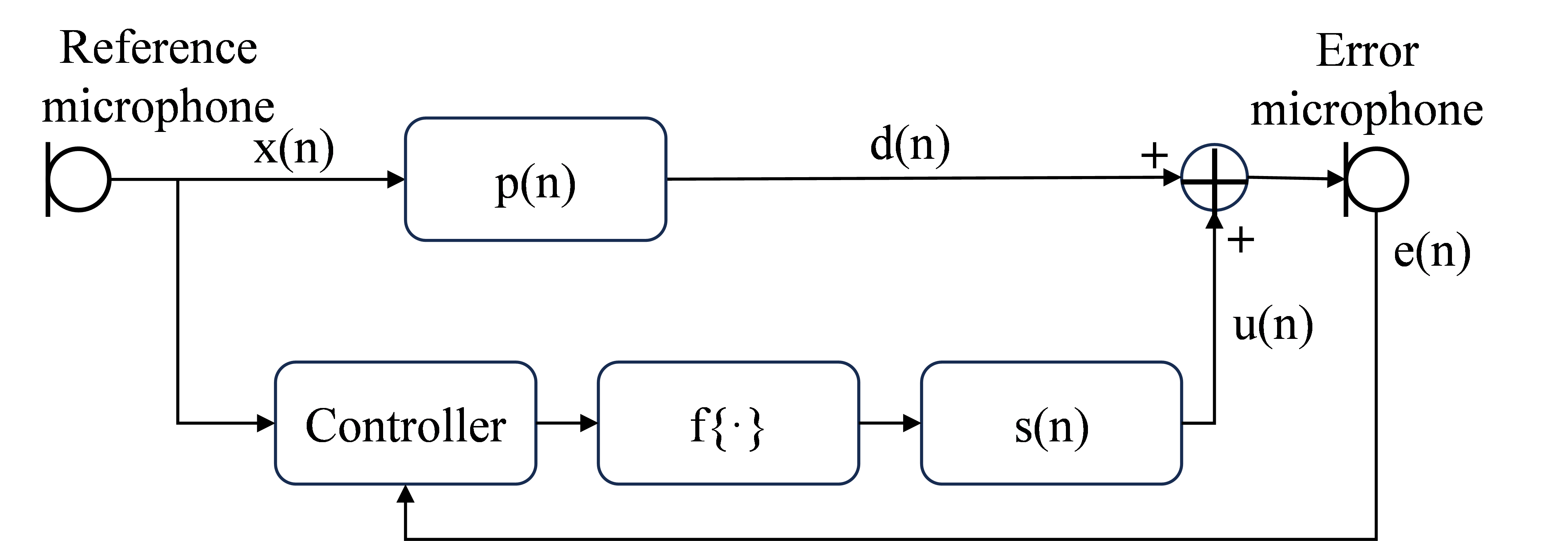}
			
	\caption{Block diagram of typical feedforward ANC system.}
	\label{fig_1}
	\vspace{-0.1cm}
\end{figure}
		
However, nonlinear distortions arise in ANC systems when actuators, such as loudspeakers, exceed their linear operating range \cite{klippel2006tutorial}. Traditional adaptive filters, designed based on a linear model, are less effective under such conditions \cite{das2004active}. To address this, researchers have proposed several nonlinear ANC algorithms. Tan et al. implemented an Volterra filtered-x LMS (VFxLMS) algorithm, which improves the ANC performance for nonlinear primary paths and reference signals \cite{tan2001adaptive}. Sahib et al. proposed a nonlinear FxLMS (NLFxLMS) algorithm based on SEF fitting of the speaker nonlinear to get copy of the linear and nonlinear models of the secondary path \cite{sahib2012nonlinear}. Ghasemi et al. improved the NLFxLMS algorithm by estimating the speaker nonlinearity through the THF function and proposed the THF-FxLMS algorithm \cite{ghasemi2016nonlinear}.
			
Recently, deep learning has been progressively utilized in the ANC field due to its powerful modeling capabilities \cite{zhang2021deep,zhang2023deep, zhang2023low,shi2022selective,chen2021secondary}. Early efforts framed ANC as a supervised learning problem and proposed end-to-end deep neural network (DNN) models for single-channel \cite{zhang2021deep} and multi-channel \cite{zhang2023deep} ANC, eliminating the need for secondary path estimation and avoiding the convergence time of traditional algorithms. More recently, the attentive recurrent network (ARN) utilizes an overlap-add method to achieve low-latency deep ANC \cite{zhang2023low}. Additionally, some researchers have proposed models that combine deep learning with traditional ANC methods. Shi et al. combined a lightweight DNN classifier with the FxLMS algorithm to obtain the selective fixed-filter active noise control (SFANC) algorithm \cite{shi2022selective}. Chen et al. proposed a DNN based secondary path-decoupled ANC (SPD-ANC) algorithm that employs time-domain CRNs to estimate the forward and reverse impulse responses (IRs) of the speaker and secondary path \cite{chen2021secondary}.

Although deep ANC provides novel solutions for addressing nonlinearity, existing comparisons may not fully reflect the potential of traditional ANC algorithms. Many studies primarily benchmark deep learning-based methods against FxLMS variants, without considering alternatives such as the Wiener filter or more advanced adaptive algorithms with optimized performance. Additionally, some evaluations employ relatively short filter lengths or unconverged traditional filters, which may underrepresent their noise reduction capabilities \cite{zhang2021deep,zhang2023deep,zhang2023low}. Furthermore, most end-to-end deep ANC approaches face challenges related to noncausal operation due to the frame-by-frame processing in DNN inference\cite{zhang2021deep,zhang2023low}, limiting their practical applicability. Under the experimental conditions of this study, we observed that existing deep ANC algorithms fail to outperform certain traditional adaptive filters with longer filter lengths and full convergence. This limitation may stem from factors such as the limited computational complexity and causality violations in certain approaches.

In this letter, we propose a fully causal time-domain ANC model by combining WaveNet \cite{van2016wavenet} and Volterra Neural Networks (VNNs) \cite{roheda2019volterra}. WaveNet has been shown to possess strong acoustic modeling capabilities \cite{van2016wavenet,rethage2018wavenet,oord2018parallel}, while the recently proposed VNNs introduces a novel way of incorporating nonlinearity through higher-order convolutions instead of traditional activation functions \cite{roheda2019volterra,roheda2024mr}. Our model is made up of a stack of dilated causal convolution layers and VNN blocks, retaining the powerful acoustic modeling capabilities of WaveNet while optimizing the model's ability to capture nonlinearity. Simulation results indicate that existing end-to-end deep learning-based ANC methods do not exhibit a clear performance advantage over fully converged traditional adaptive filters with longer filter lengths. In contrast, the proposed method consistently outperforms both other DNN-based ANC algorithms and traditional approaches across different metrics.

\section{Overview of Current ANC Algorithms} \label{2}
		
The signal processing block diagram of the typical feedforward ANC system is illustrated in Fig. \ref{fig_1}. 
Specifically, the reference signal \( x(n) \) generates the noise \( d(n) \) through the primary path \( p(n) \), while the reference signal \( x(n) \) passes through the control filter \( w(n) \) to produce \( y(n) \). Then, the control signal \(y(n)\) passes through the secondary path consisting of the loudspeaker and the acoustic secondary path \(s(n)\), resulting in the anti-noise signal \(u(n)\), which superimposes with the noise signal \(d(n)\) to form the error signal \(e(n)\), expressed as
\begin{equation}
	\label{equation1}
	\begin{aligned}
		e(n) &= d(n) + u(n) \\
		&= p(n) * x(n) + s(n) * f\{w(n) * x(n)\}\\
	\end{aligned}
\end{equation}
where \(*\) denotes linear convolution and \(f\{\cdot\}\) denotes the loudspeaker response function.
For the linear TD-FXLMS algorithm, the filter coefficients are updated using the gradient descent method as follows:
\begin{equation}
	\label{equation2}
	\w(n+1) = \w(n) - \mu e(n)\rr(n) \\
\end{equation}
where \(\w(n)\) denotes weighted vector and \(\rr(n)\) denotes the filtered reference signal vector.
		
Deep ANC replaces the control filter in the adaptive filter with a DNN model, eliminating the need for secondary path estimation during training and avoiding secondary path modeling errors. Currently, most frequency-domain deep ANC algorithms suffer from high latency, and their causal violation in the prediction methods results in suboptimal performance when applied to complex systems.
				
\section{WaveNet-VNNs Model}\label{3}
		
\subsection{WaveNet}
WaveNet is a generative convolutional neural network (CNN) model that operates directly on the raw audio waveform. It constructs the complete audio sequence by progressively generating each sample point of the audio signal. The goal of WaveNet is to model the joint probability distribution of the entire sequence as follows:
\begin{equation}
	\label{equation7}
	p\left( {\bf x}\right) = \prod_{t=1}^T p\left( x_t\mid x_1,\ldots,x_{t-1}\right)
\end{equation}
		
\begin{figure*}[!t]
	\centering
	\vspace{-0.2cm}
	\includegraphics[width=0.8\textwidth]{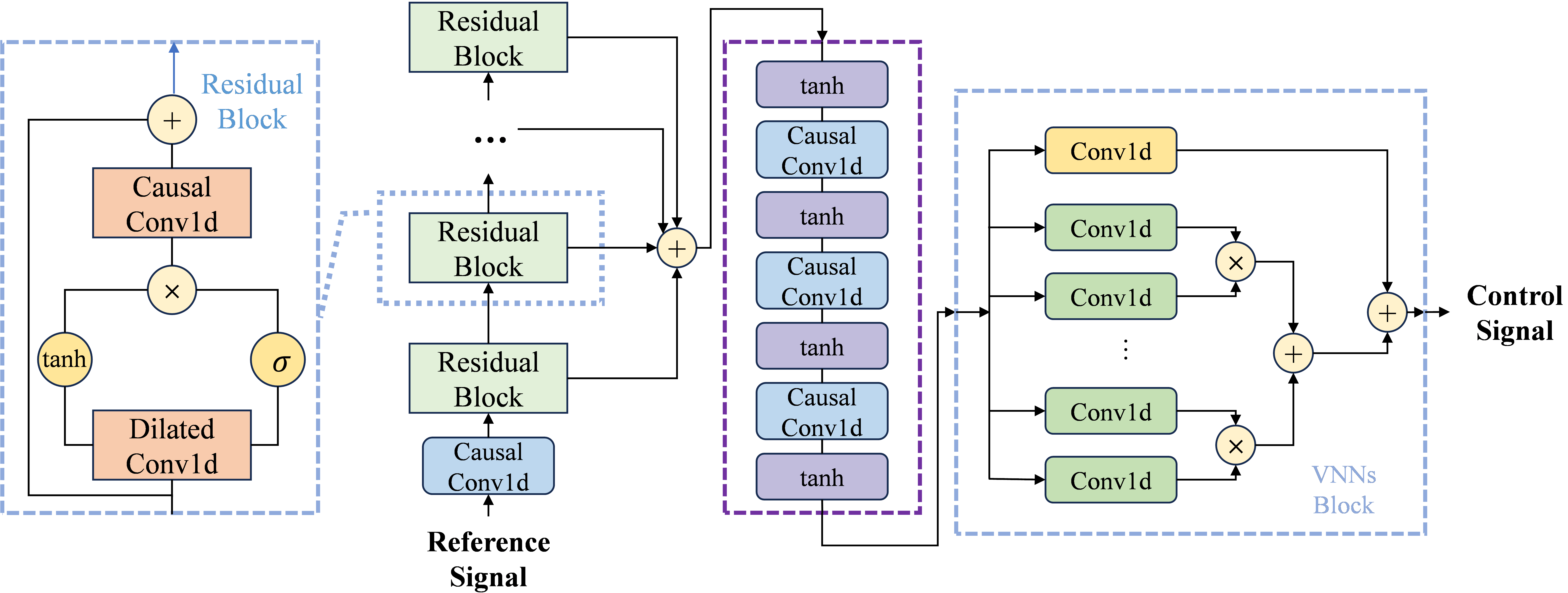}
	\caption{\centering{ The structure of the WaveNet-VNNs for ANC.}}
	\vspace{-0.2cm}
	\label{fig_4}
\end{figure*}
		
One of its core components is the causal dilated convolution. This convolution structure ensures that each output sample depends only on the current and previous input samples, thereby satisfying the strict causal requirements. By introducing dilated convolution, WaveNet significantly expands the receptive field while maintaining parameter efficiency, allowing it to capture long-range dependencies in the audio signal. To further enhance the model's nonlinear expression capability, WaveNet employs the following gated unit:
\begin{equation}
	\label{equation8}
	\mathbf{z} = \tanh \left( W_{f, k} * \mathbf{x} \right) \odot \sigma \left( W_{g, k} * \mathbf{x} \right)
\end{equation}
where $\ast$ denotes the convolution operation, $\odot$ represents the element-wise multiplication operator, $\sigma(\cdot)$ is the sigmoid function, $k$ is the layer index, $f$ and $g$ correspond to the filter and gate, respectively, and $W$ is a learnable convolution filter. The stacking of residual blocks and skip connections accelerates the model's convergence speed.

\subsection{Volterra Neural Networks}
Volterra Neural Networks is a network architecture inspired by the Volterra filter. The goal of VNNs is to efficiently model nonlinear dynamic systems by approximating the kernel functions of the Volterra series using neural networks \cite{roheda2019volterra}. Instead of traditional activation functions, it incorporates nonlinearity into the system response function through higher-order convolutions. A typical VNNs block consists of multiple layers of convolutions, where each layer represents interactions of different orders between the input signals.

\subsection{WaveNet-VNNs for ANC}
		
As shown in Fig. \ref{fig_4}, our model is composed of a series of WaveNet and VNNs blocks connected in sequence. The model directly processes the time-domain signal without involving any frequency-domain operations. The input is the reference signal, and the output is the control signal. The reference signal is first passed through a causal convolution layer, followed by a series of residual blocks that include dilated causal convolutions, gated units, and causal convolutions. The results after the skip connections are processed through four tanh activation functions and three causal convolution layers. Finally, to enhance the network's nonlinear modeling capability, the signal is passed through a VNNs block to obtain the final control signal. The VNNs block used here consists of first-order and second-order convolutions, specifically including one causal convolution layer and four sets of second-order convolution layers. All convolution layers used in the model are strictly causal.

\vspace{-0.05cm}
\section{Experiments }\label{4}

\subsection{Experimental Setting}
In the model training, we use 26 hours noise data from the DEMAND \cite{thiemann2013diverse} and MS-SNSD \cite{reddy2019scalable} datasets sliced into 3-second segments. Factory1, babble, engine and factory2 noise from NOISEX-92 \cite{varga1993assessment} dataset are used as the test set, which are not accessible during the training process. The training and test data are both subsampled to 16kHz and normalized.
		
The simulations are conducted in a rectangular room with dimensions of 3 m × 4 m × 2 m (width × length × height) based on the settings provided in \cite{zhang2023low}. The reference microphone is located at the position (1.5, 1, 1) m, the error microphone at (1.5, 3, 1) m, and the control source at (1.5, 2.5, 1) m. The primary and secondary paths are simulated as room impulse responses (RIRs) by employing the image method, as described in \cite{allen1979image}. The reverberation time (T60 s) is set to 0.2 s, and the primary and secondary paths are set to 512-length. According to \cite{zhang2021deep,zhang2023deep,zhang2023low,chen2021secondary}, we select the scaled error function (SEF) \cite{klippel2006tutorial} as the nonlinear function for the loudspeaker:
\begin{equation}
	\label{equatio9}
	f_{\text{SEF}}(y) = \int_0^y e^{-\frac{x^2}{2\eta^2}} \, dx
\end{equation}
where \(y\) is the input to the speaker, and \(\eta^2 \) denotes the strength of the nonlinearity. We train and test our model under three different nonlinear conditions with \(\eta^2 = \infty \), \(\eta^2 = 0.5 \) and \(\eta^2 = 0.1 \) respectively.
		
The proposed model consists of 30 residual layers, where the dilation factor in each layer follows a progression from 1 to 2, 4, 8, ..., up to 512. Further parameter details can be found in the open-source code.
		
The loss function used in the model training is the average of NMSE and A-weighted decibel (dBA) \cite{nilsson2007weighted} and the model is trained for 30 epochs using the Adam optimizer \cite{reddi2019convergence}.

\begin{table*}[!t]
	\vspace{-0.1cm} 
	\caption{Test NMSE (dB) and A-weighted decibel (dBA) of various traditional algorithms and the deep ANC model\label{tab:table1}}
	\centering
			
	\resizebox{\textwidth}{!}{
				
\begin{tabular}{c|c|c c c c|c c c c|c c c c}
	\toprule[0.5mm]
	\multicolumn{2}{c|}{Nonlinearity}
	&\multicolumn{4}{c|}{Linear($\eta^2 = \infty$)}
	&\multicolumn{4}{c|}{Nonlinear($\eta^2 = 0.5$)}
	&\multicolumn{4}{c}{Nonlinear($\eta^2 = 0.1$)} \\
	\cline{1-14}
	\multicolumn{2}{c|}{Noise}
	&\multicolumn{2}{c|}{Babble}
	&\multicolumn{2}{c|}{Factory1}
	&\multicolumn{2}{c|}{Babble}
	&\multicolumn{2}{c|}{Factory1}
	&\multicolumn{2}{c|}{Babble}
	&\multicolumn{2}{c}{Factory1} \\
	\cline{1-14}
					
	\multicolumn{2}{c|}{Loss}  &dBA  & NMSE &dBA & NMSE &dBA & NMSE &dBA & NMSE &dBA  & NMSE &dBA  & NMSE\\
	\hline
					
	\multicolumn{2}{c|}{SPD-ANC}  &-5.17  & -7.81 &-4.08 & -9.35 &-5.17 & -7.56 &-4.08 & -9.34 &-5.14  & -6.84 &-4.06  & -9.30\\
	\hline
					
	\multicolumn{2}{c|}{CRN}  &-14.31  & -14.80 &-11.81 & -13.30 &-14.31 & -14.40 &-11.80 & -13.25 &-14.14  & -12.12 &-11.66  & -12.94\\
	\hline
					
	\multicolumn{2}{c|}{THF-FxLMS(512)}  &--  & -- &-- & -- &-8.51 & -14.21 &-11.99 & -12.34 &-8.41  & -11.53 &-11.87  & -12.30\\
					
	\multicolumn{2}{c|}{TD-FxLMS(512)}  &-8.52  & -14.47 &-12.00 & -12.35 &-8.51 & -13.86 &-11.99 & -12.35 &-8.44  & -10.33 &-11.89  & -12.30\\
					
	\multicolumn{2}{c|}{FD-FxNLMS(512)}  &-12.46  & -14.95 &-12.00 & -11.51 &-12.46 & -14.38 &-11.99 & -11.52 &-12.39 & -12.15 &-11.88  & -11.50\\

	\multicolumn{2}{c|}{ODW-FDFeLMS(512)}  &-12.56  & -14.10 &-11.98 & -9.41 &-12.55 &-13.61  &-11.97 & -9.41 &-12.44  & -11.41 &-11.80  & -9.38\\

	\multicolumn{2}{c|}{Wiener(512)}  &-13.09  & -17.18 &-12.34 & -12.33 &-13.08 &-16.56  &-12.33 & -12.32 &-12.96  & -13.29 &-12.20  & -12.22\\
	\hline
					
	\multicolumn{2}{c|}{TD-FxLMS(2048)}  &-7.43  & -14.65 & -15.52 &-20.50 &-7.41 & -7.58 &-15.47 & -20.42 &-7.32  & -9.05 &-14.96  & -19.65\\
					
	\multicolumn{2}{c|}{FD-FxNLMS(2048)}  &-28.51  & -27.67 &-25.02 & -26.08 &-28.19 & -23.34 &-24.77 & -25.95 &-24.90  & -15.28 &-21.69  & -23.28\\
					
	\multicolumn{2}{c|}{ODW-FDFeLMS(2048)}  &-29.00  & -33.65 &-26.64 & -27.92 &-28.56 & -26.24 &-26.21 & -27.63 &-24.66  & -15.52 &-21.92  & -23.87\\
					
	\multicolumn{2}{c|}{Wiener(2048)}  &-30.79  & -36.00 &-27.02 & -29.30 &-30.09 & -25.00 &-26.50 & -28.72 &-25.39  & -15.27 &-22.31  & -23.42\\
	\hline
					
	\multicolumn{2}{c|}{WaveNet-VNNs}  &\textbf{-34.48}  & \textbf{-41.04} &\textbf{-29.58} & \textbf{-35.44} &\textbf{-33.23} & \textbf{-35.02} &\textbf{-28.84} & \textbf{-34.06} &\textbf{-30.50}  & \textbf{-17.22} &\textbf{-26.38}  & \textbf{-27.94}\\
	\bottomrule[0.5mm]
	\end{tabular}}
	\vspace{-0.1cm} 
\end{table*}

\vspace{-0.05cm}
\subsection{Baseline Methods}
We campare our proposed model with TD-FxLMS \cite{elliott1987multiple}, tangential hyperbolic function based FxLMS(THF-FxLMS) \cite{ghasemi2016nonlinear}, FD-FxNLMS \cite{yang2018frequency}, ODW-FDFeLMS \cite{lian2024online}, Wiener filter \cite{elliott2000signal}, SPD-ANC \cite{chen2021secondary} and convolutional recurrent network (CRN) \cite{zhang2021deep} in three different nonlinear conditions. 
		
In particular, TD-FxLMS, FD-FXLMS, ODW-FeLMS, and Wiener filter are evaluated with control filter lengths of both 512 and 2048 points. In the linear case, the noise reduction performance of the Wiener filter can theoretically improve indefinitely as the number of control filter taps increases \cite{haykin2014adaptive}. However, we do not choose filter lengths greater than 2048, as the improvement in performance is relatively marginal for linear cases, and there is almost no improvement in nonlinear situations. To achieve optimal solutions with traditional algorithms, we utilized the precise secondary path directly. Note that all subsequent comparisons are based on fully converged results to ensure accuracy. Additionally, for the Wiener filter, the same dataset is employed for both training and testing.
		
The optimal step sizes are employed in all algorithms to ensure the best convergence performance, and these values are determined through a trial-and-error approach.

\vspace{-0.05cm}
\subsection{Experimental Results}
		
\begin{figure}[!t]
	\centering
	\includegraphics[width=\linewidth]{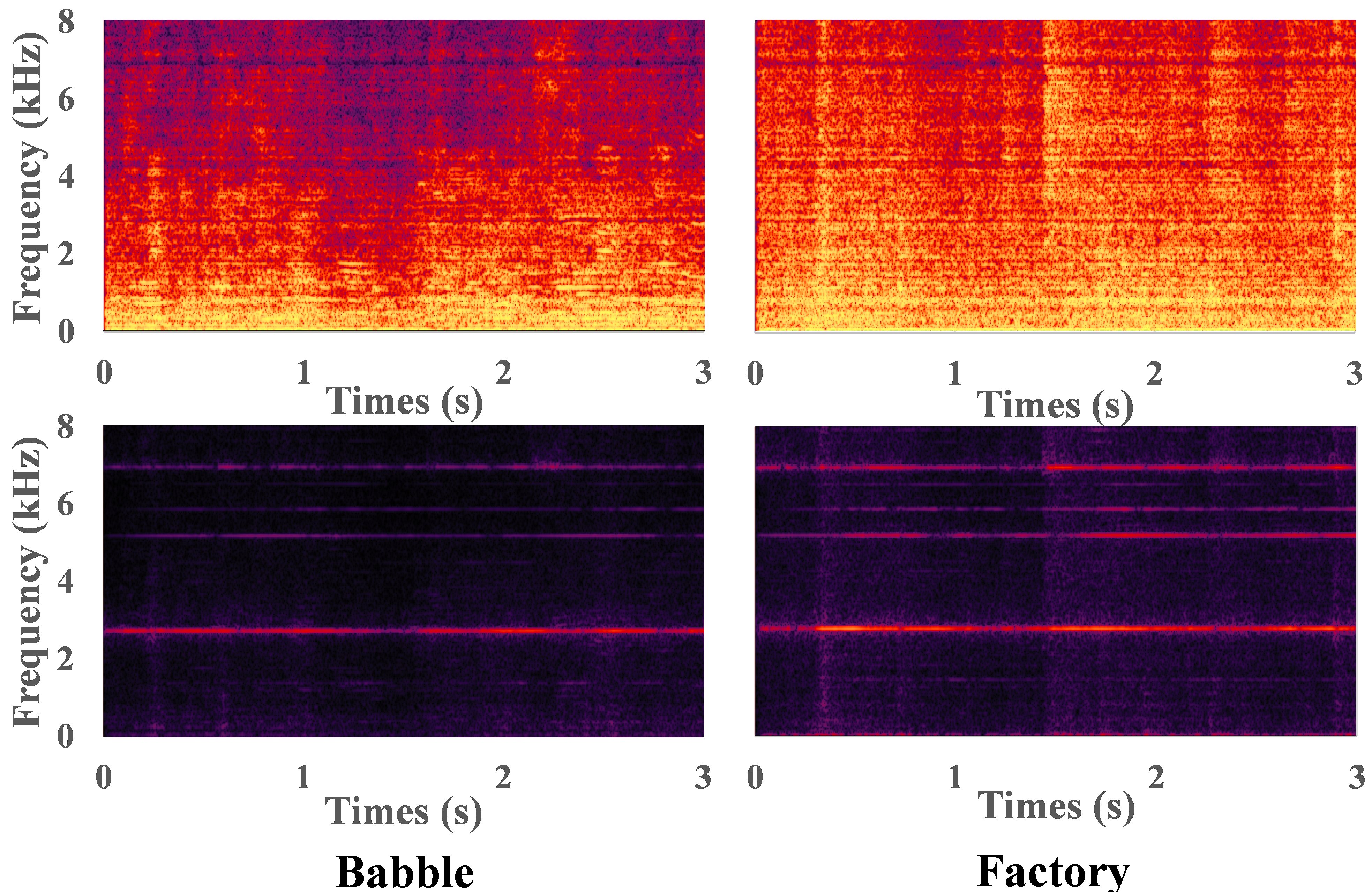}
	\caption{Typical spectrograms of test results for two types of noise. The first row shows the results with ANC off, and the second row shows the results with ANC on.}
	\label{fig_5}
\end{figure}
		
We evaluate the ANC algorithms under both linear (\(\eta^2 = \infty \)), and nonlinear conditions (\(\eta^2 = 0.5, 0.1\)) using noise that is unseen during the training process. The results are shown in Table \ref{tab:table1}. The test signal used is the complete noise segment from the Noisex-92 database, which is normalized without any truncation. The numbers in parentheses in the Table \ref{tab:table1} represent the number of taps in the control filter. 
		
It can be observed that the fully converged traditional algorithms outperform the two baseline Deep-ANC algorithms under both linear and nonlinear conditions. The relatively poor performance of SPD-ANC may be attributed to the use of more complex forward and reverse impulse responses of the speaker and secondary path in the time-domain CRN fitting, as opposed to the simple model with only a few points used in the original paper \cite{chen2021secondary}. The resulting modeling errors due to the complexity of the secondary paths adversely affect its performance.
		
\begin{table}[!t]
			
	\caption{ANC performance of wiener filter (2048) and WaveNet-VNNs for engine and factory2 noises\label{tab:table2}}
	\centering
	\renewcommand{\arraystretch}{1.2}
	\begin{tabular}{c|c|c c |c c}
		\toprule[0.5mm]
		\multicolumn{2}{c|}{Method}
		&\multicolumn{2}{c|}{Wiener(2048)}
		&\multicolumn{2}{c}{WaveNet-VNNs}
		\\
		\cline{1-6}
		Noise  &Nonlinearity  &dBA & NMSE &dBA & NMSE \\
		\cline{1-6}

		\multirow{3}{*}{Engine} & $\eta^2 = \infty$  &-34.21 & -33.63 &-35.46 & -35.25 \\ 
		& $\eta^2 = 0.5$ &-28.79 & -28.40 &-31.06 & -29.76\\ 
		& $\eta^2 = 0.1$ &-17.95 & -17.42 &-22.16 & -19.93\\
		\cline{1-6}
				
		\multirow{3}{*}{Factory2} & $\eta^2 = \infty$  &-29.40 & -30.46 &-33.25 & -46.07 \\ 
		& $\eta^2 = 0.5$ &-25.31 & -27.20 &-30.53 & -37.95\\ 
		& $\eta^2 = 0.1$ &-15.49 & -17.35 &-24.01 & -23.43\\
		\bottomrule[0.5mm]
	\end{tabular}
	\vspace{-0.2cm} 
\end{table}
		
Among the traditional algorithms, TD-FxLMS exhibits the weakest performance, whereas FD-FxNLMS significantly improves stability through block updates, leading to a substantial performance gain over TD-FxLMS. ODW-FDFeLMS achieves even better results, primarily due to its multi-frame update mechanism, which enhances stability and improves robustness against large-amplitude impulse noise. The Wiener filter demonstrates the best performance in most cases, as it provides the statistically optimal solution in the least-squares sense.
		
It is important to note that the performance of all algorithms significantly deteriorates under nonlinear conditions for the babble signal due to the presence of a high-amplitude pulse in the error signal generated through the primary path. It can be observed that the performance of most traditional algorithms improves as the number of taps in the control filter increases, except for the TD-FxLMS algorithm for the babble signal. This is because the normalized babble signal used in the test generates an error signal with a large-amplitude pulse through the primary path. Since TD-FxLMS updates point by point, its stability worsens at 2048 taps.
		
Furthermore, as the degree of nonlinearity increases, all methods exhibit performance degradation, with the most pronounced effects observed in traditional algorithms, including the Wiener filter. To further assess our model, we compare the Wiener filter and WaveNet-VNNs with two additional noise types (engine and factory2), as shown in Table \ref{tab:table2}. Our model consistently achieves superior performance across all cases, surpassing the Wiener filter in every scenario, with particularly notable improvements in nonlinear conditions.
		
To better illustrate the performance, we present typical spectrograms of two types of noise before and after ANC using our model in Fig. \ref{fig_5}. It can be observed that our model performs well in terms of noise reduction level across a wide frequency range. Several abnormal frequency points is caused by the notch characteristics at corresponding frequency points in the generated room secondary path.

\section{Conclusion}
		
In this letter, we propose a deep ANC model by combining WaveNet and VNNs. Simulation results show that our model outperforms traditional algorithms, including the Wiener filter, as well as some of the currently proposed DeepANC algorithms, in terms of both NMSE and dBA metrics. Compared to the best-performing Wiener-2048, our model achieves an average improvement of 3.85 dBA and 5.91 dB (NMSE) across four test noise types under three different nonlinear conditions. Furthermore, our model is fully causal and introduces no delay, assuming no computational constraints. The scope of this study is limited to single-channel simulation scenarios, and future work will focus on simulations in more complex multi-channel real-world environments. 

\newpage

\bibliographystyle{IEEEtran}
\bibliography{reference}

\end{document}